# On the Informational Aspects of Interfering Quantum States

**Yu.I. Bogdanov*, A.Yu. Bogdanov*, S.A. Nuyanzin**, A.K. Gavrichenko***

*Institute of Physics and Technology, Russian Academy of Sciences*[1]*
*Moscow Institute of Electronic Technology (Technical University)***

In present work we study informational measures for the problem of interference of quantum particles. We demonstrate that diffraction picture in the far field, which is given by probability density of particle momentum distribution, represents a mixture of probability densities of corresponding Schmidt modes, while the number of modes is equal to the number of slits at the screen. Also, for the first time we introduce informational measures to study the quality of interference picture and analyze the relation between visibility of interference picture and Schmidt number. Furthermore, we consider interference aspects of the problem of a quantum particle tunneling between two potential wells. This framework is applied to describing various isotopic modifications of ammonia molecule. Finally, we calculate limits on the maximum possible degree of entanglement between quantum system and its environment, which is imposed by measurements.

## 1. Introduction

Interference of quantum states is one of key concepts in quantum informatics, which ensures superiority of quantum computers over its classical counterparties for certain types of problems [1-3]. At the same time the loss of interference of quantum states (decoherence) remains one of the main practicality issues in building a full-scale quantum computer.

The traditional approach to the concept of interference of quantum states, proposed by Feynman, is reduced to considering distinguishable and undistinguishable alternatives [4]. It implies that distinguishable alternatives interfere, contrary to undistinguishable ones. The problem of this approach is that it considers only two limiting cases: interfering and non-interfering alternatives. Therefore, in such framework it is impossible to describe a more common case of partially distinguishable and undistinguishable interfering alternatives.

There is a long history for the studies of interaction between a quantum system and its environment. Bohr, during his debate with Einstein about principles of quantum mechanics, illustrated his complementarity principle by offering a hypothetical theoretical construction of an optical interferometer with a versatile screen with slits [5]. The interaction between a particle and the screen allowed one to figure out which slit the particle went through by carrying out measurements on the screen. Measurements of the particle’s path (i.e. obtaining "which way" information) lead to disappearance of interference picture.

Today, using modern experimental technology Bohr's thought experiment was realized in practice to certain extent [6-8].

For instance, in work [6] photon pairs produced during the process of spontaneous parametric down-conversion were used. In order to observe photon interference it was necessary that no measurements of "which way" type were carried out. If, on the other hand, the second photon was used to register the first photon’s path, then the interference picture disappeared.

Furthermore, in work [7] a two-beam electron interferometer located in the plane of two-dimensional electron gas with very high mobility was used. The two electron paths from emitter to collector in the interferometer were defined by appropriate "slits" for electrons. One of them was made like a coherent quantum dot. The quantum dot is a trap which captures electrons for a relatively long time like resonance delay line that facilitated the electron registration. Near the quantum dot, though electrically isolated from it, there was located a quantum point contact, which served as the "which way" detector. The corresponding detection process led to dephasing and loss of two-beam Aharonov- Bohm interference.

[1] E-mail: bogdanov@ftian.ru

Finally, in work [8] an atomic two-beam Ramsey interferometer was used. There microwave impulses played the role of beam-splitters for atomic states. One of these impulses presented by itself a coherent state of radiation in a cavity with a relatively low photon number. The visibility of the interference picture improved with the growth of the average number of photons in the coherent state. It was an illustration of transition from quantum consideration to classical description.

Note that theoretical approaches to explaining erosion of interference picture influenced by measurements and environment, have differed throughout the whole history of quantum mechanics.

Therefore in his arguments Bohr appealed to the principle of complementarity based on numerical estimations, following from uncertainty relations for the variables "coordinate-momentum". However, in work [9] (see also [10]) it is argued that the erosion of the interference picture is not related to uncertainty relations and is instead caused by quantum correlations appearing due to entanglement between interfering particle with measurement device. See the book by Mensky [11] for a throughout description of the debate, as well as analysis of "which-way" experiments. Recent work [12] is also dedicated to discussing these questions.

The physical cause for distinguishability of quantum states and loss of coherence is related to the entanglement phenomenon between quantum state and environment. At the same time, so far, quantum correlations have only been able to provide qualitative rather than quantitative estimates of informational aspects of the problem.

Mathematically, entanglement may be described by so-called Schmidt decomposition [1, 3]. In works [13-15] Schmidt decomposition is performed for physical systems with continuous variables and the concept of Schmidt information, defining quantitative measures of entanglement, is introduced.

The purpose of this work is to construct explicit mathematical models for entangled states of quantum system with environment and analyze them in the framework of Schmidt decomposition.

This paper is organized as follows.

In section 2 two-slit Young interference is briefly discussed. The Young's experiment gives fundamental illustration for the phenomenon of quantum superposition. From the informational viewpoint the Young's experiment is characterized by mutually complementary data obtained near the screen for coordinate measurements and in the far field for momentum measurements correspondingly.

In section 3 a simple analytical model describing an entangled joint state of interfering and detecting particles is offered. The considered model gives a quantitative description for the process of interference picture degrading with the growth of interaction between a particle and the detector.

In section 4 for the first time we provide an analysis of problem "which way" using Schmidt decomposition for joint entangled state of interfering particle and detector. Explicit expressions for particle and detector modes, as well as Schmidt decomposition weights are obtained.

In sections 5 and 6 the considered model is generalized for the case of interference on the screen with arbitrary number of slits.

In section 7 for the first time we develop a quantum information approach to describe classical coherence. We propose a formal model that explains the phenomenon of loss of coherence as consequence of entanglement of optical system with auxiliary quantum system. It's shown that within the developed approach there is a universal relation between visibility of interference picture and Schmidt number.

In sections 8 and 9 the interference approach is applied to considering the problem of a quantum particle tunneling between two potential wells. Similarities between this problem and two-slit interference are noted.

In section 10, the model from section 8 is applied to studying ammonia molecule and its isotopic modifications. Close agreements between theory and experiment are observed.

In section 11, we consider interference control from the viewpoint of adequacy and completeness of quantum measurements. It is shown that measurements impose some restrictions on the possible form of quantum states.

In section 12, the work summary is given.

## 2. Two-slit interference - Young's experiment as fundamental illustration of quantum superposition

Two-slit interference, observed by Young, is the simplest and most important example of interference. It is important to understand the fundamentals of quantum mechanics [4, 16] – in Feynman's words it "is devised to cover all the mysteries of quantum mechanics, and familiarize one with all the paradoxes, secrets and peculiarities of the nature. In quantum mechanics it's always possible to explain any phenomenon saying: "Do you remember our experiment with two slits? There is the same story here"" (translation from Russian edition of [16]).

In our problem, the variables are continuous (coordinate and momentum). Still, it may be described by means of the most simple two-level quantum system. In this representation, a quantum particle interfering on two slits is the simplest realization of a quantum bit (qubit). The corresponding mathematical model may be applied not only to slit interference, but to many other interesting physical situations (for example, see section 8 - quantum particle tunneling between two potential wells).

In this section, we shall consider a quantum particle interfering on two Gauss-type slits. We shall also provide exact formulas for normalization coefficients, which are usually omitted in other works. In the end of the section some paradoxes and contradictions of the one- particle quantum model will be shown.

Let's set probability distribution of each slit:

$$P(x)=\frac{1}{\sqrt{2\pi}\sigma_x}\exp\left(-\frac{(x-a)^2}{2\sigma_x^2}\right)$$

This formula describes normal distribution with mean $a$ and variance $\sigma_x^2$.

We shall suppose that this distribution was generated by a particle with quantum state:

$$\psi(x)=\sqrt{P(x)}=\left(\frac{1}{\sqrt{2\pi}\sigma_x}\right)^{1/2}\exp\left(-\frac{(x-a)^2}{4\sigma_x^2}\right)$$

In momentum representation the wave function is derived by a Fourier transformation (we suppose that $\hbar=1$):

$$\tilde{\psi}(p_x)=\frac{1}{\sqrt{2\pi}}\int\psi(x)\exp(-ip_x x)dx=\left(\frac{1}{2\pi}\right)^{1/4}\sqrt{2\sigma_x}\exp\left(-\sigma_x^2p_x^2\right)\exp(-ip_x a)$$

The appropriate probability distribution is:

$$\tilde{P}(p_x)=\left|\tilde{\psi}(p_x)\right|^2=\left(\frac{1}{2\pi}\right)^{1/2}2\sigma_x\exp\left(-2\sigma_x^2p_x^2\right) \qquad (1)$$

This distribution is also Gaussian with variance $\sigma_p^2 = \frac{1}{4\sigma_x^2}$ (in according with uncertainty relation).

Now let us consider two Gauss-type slits centered at points $+a$ and $-a$ accordingly (that is the distance between slit centers equals $2a$ ).

The corresponding wave function is given by a symmetric superposition:

$$\psi(x) = C\left(\frac{1}{\sqrt{2\pi}2\sigma_x}\right)^{1/2}\left[\exp\left(-\frac{(x-a)^2}{4\sigma_x^2}\right) + \exp\left(-\frac{(x+a)^2}{4\sigma_x^2}\right)\right],$$

where $C$ - is normalization constant:

$$C^2 = \frac{1}{1+\exp\left(-\frac{a^2}{2\sigma_x^2}\right)}$$

It's evident that $C \approx 1$, if $a >> \sigma_x$ (in case of slits distinctly separated from each other). In momentum representation the two-slit wave function is:

$$\tilde{\psi}(p_x) = 2C\left(\frac{1}{2\pi}\right)^{1/4}\sqrt{\sigma_x}\exp\left(-\sigma_x^2 p_x^2\right)\cos(p_x a)$$

The interference picture is given by probability distribution in momentum representation:

$$\tilde{P}(p_x) = |\tilde{\psi}(p_x)|^2 = 4C^2\left(\frac{1}{2\pi}\right)^{1/2}\sigma_x\exp\left(-2\sigma_x^2 p_x^2\right)\cos^2(p_x a) \qquad (2)$$

Two-slit interference picture is given in Fig.1

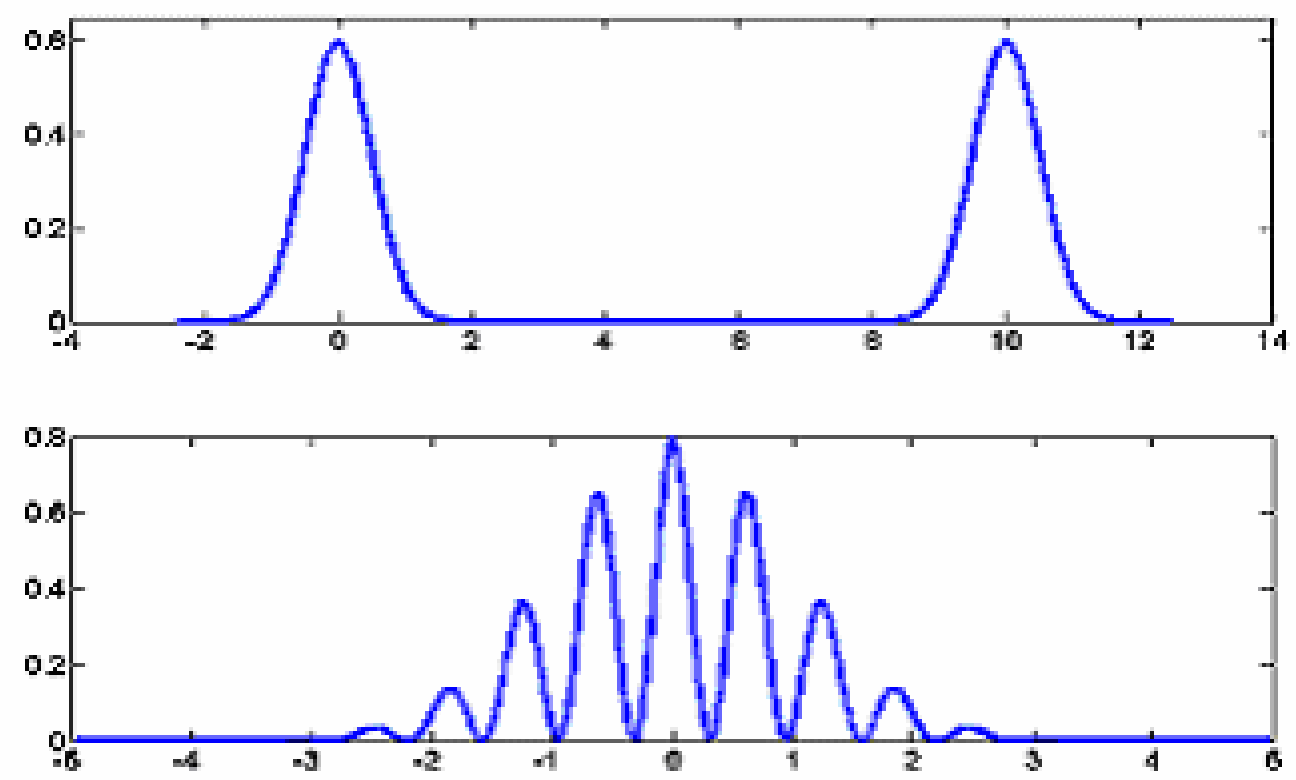

Figure 1. Two-slit interference picture: top - coordinate distribution, bottom – momentum distribution

Clearly, if only one slit is opened, then intensity in the far field is given by Gauss distribution (formula (1)); if both slits are there opened, then intensity has the form of a typical interference picture (2).

A seemingly reasonable question arises - how does the particle going through one of the slits "know" about existence of the second slit (and how does it "feel" the second slit's state (open, close))? However, such wording of the question implies an explicit possibility to separate the state of a single particle from the states of all other particles in the universe. In quantum mechanics it is not the case. It is impossible to separate a single-particle wave function from a multi-particle entangled wave function. Therefore, it follows that multi-particle entangled states should be considered.

**3. Expansion of quantum system by addition of new variables. Entanglement with environment as the way of quantum superposition limitation.**

The difficulties and paradoxes described above would disappear if the particle's environment was taken into consideration.

The quantum system presented above consists of a single particle. It was described by wave function $\psi(x)$ in coordinate or $\tilde{\psi}(p_x)$ in momentum representation. Now let us consider a more complex two-dimension function $\psi(x,\xi)$, where $x$ shall be interpreted as a coordinate of the particle and $\xi$ - as a coordinate of some another (detecting) particle, which interacts with the original particle. The latter parameter may also be described as one registered by some detector monitoring activity of the original particle. (for example, parameter $\xi$ can represent a photon scattered by a particle going through diaphragm).

Any quantum measurement, irrelevant of its realization, may be described in the following shortened form: factorized state of particle and detector at the input is transformed to a joint entangled state at the output. Schematically it can be presented as:

$$\psi_{in}^{x}(x)\otimes\psi_{in}^{\xi}(\xi)\underset{measurement}{\rightarrow}\psi_{out}^{x,\xi}(x,\xi)$$

First, let us consider a model given by superposition of two two-dimensional Gauss probability distributions:

$$\psi(x,\xi)=\frac{C}{\sqrt{2}}\frac{1}{\sqrt{2\pi\sigma_x\sigma_\xi}}\left[\exp\left(-\frac{(x-a)^2}{4\sigma_x^2}-\frac{(\xi-b)^2}{4\sigma_\xi^2}\right)+\exp\left(-\frac{(x+a)^2}{4\sigma_x^2}-\frac{(\xi+b)^2}{4\sigma_\xi^2}\right)\right] \quad (3)$$

Here the detector is described by two Gauss distributions with expectations $\pm b$ and variance $\sigma_\xi^2$. Normalization constant is given by:

$$C^2=\frac{1}{1+\exp\left(-\frac{1}{2}\left(\frac{a^2}{\sigma_x^2}+\frac{b^2}{\sigma_\xi^2}\right)\right)}$$

Let's consider the structure of wave function. As a particle passes through the first slit at point $x=+a$, the detector measures parameter $\xi=+b$. Similarly, for the second slit, $\xi=-b$. So the measurements by detector correlate with position of the particle. In the framework of quantum mechanics, this state is entangled: coordinates $x$ and $\xi$ correlate (are entangled) with each other.

The simplest physical realization of quantum state (3) is as follows. Let $x$ and $y$ be horizontal and vertical coordinates of the particle correspondingly. Then the parameter $y$ may play the role of parameter $\xi$ itself. In our considered case coordinates $x$ and $y$ will correlate with one another. Measurement of coordinate $y$ will bring some information about coordinate $x$, resulting in partial or full loss of interference. Also, note that in experiments of type [6], parameters $x$ and $\xi$ may be represented by coordinates of twin-photons, emitted as the result of spontaneous parametric down-conversion.

Evidently the correlation between $x$ and $\xi$ is very high for $b >> \sigma_\xi$. In this case the interference picture disappears as confirmed by calculations below.

Two-dimensional wave function in momentum representation is given by:

$$\tilde{\psi}(p_x, p_\xi) = \frac{2C}{\sqrt{2}}\left(\frac{1}{2\pi}\right)^{1/2} \sqrt{2\sigma_x}\sqrt{2\sigma_\xi}\exp\left(-\sigma_x^2 p_x^2\right)\exp\left(-\sigma_\xi^2 p_\xi^2\right)\cos\left(p_x a + p_\xi b\right)$$

Two-dimensional (two-particle) probability distribution in momentum representation is:

$$\tilde{P}(p_x, p_\xi) = \left|\tilde{\psi}(p_x, p_\xi)\right|^2$$

We shall derive the marginal distribution, which corresponds to particle's momentum, by integrating by the probability distribution by detector's momentum.

$$\tilde{P}_x(p_x) = \int \tilde{P}(p_x, p_\xi)dp_\xi$$

As a result:

$$\tilde{P}_x(p_x) = \frac{1}{1 + \exp\left(-\frac{1}{2}\left(\frac{a^2}{\sigma_x^2} + \frac{b^2}{\sigma_\xi^2}\right)\right)}\sqrt{\frac{2}{\pi}}\sigma_x \exp\left(-2\sigma_x^2 p_x^2\right)\left[1 + \exp\left(-\frac{b^2}{2\sigma_\xi^2}\right)\cos(2p_x a)\right]$$

For $b = 0$ the interference picture coincides with the ideal one-dimensional case described in the previous section.

Similarly, two-dimensional coordinate distribution:

$$P(x,\xi) = \left|\psi(x,\xi)\right|^2$$

Marginal distribution of coordinate of the particle:

$$P_x(x) = \int P(x,\xi)d\xi$$

Then:

$$P_x(x) = \frac{1}{2\left[1 + \exp\left(-\frac{1}{2}\left(\frac{a^2}{\sigma_x^2} + \frac{b^2}{\sigma_\xi^2}\right)\right)\right]}\frac{1}{\sqrt{2\pi}\sigma_x}\times$$

$$\times\left[\exp\left(-\frac{(x-a)^2}{2\sigma_x^2}\right) + \exp\left(-\frac{(x+a)^2}{2\sigma_x^2}\right) + 2\exp\left(-\frac{1}{2}\left(\frac{a^2}{\sigma_x^2} + \frac{b^2}{\sigma_\xi^2}\right)\right)\exp\left(-\frac{x^2}{2\sigma_x^2}\right)\right]$$

If the slits are distinctly separated from each other ($a >> \sigma_x$), then the last summand describing their overlapping is negligibly small.

These expressions demonstrate the following – addition of a new variable $\xi$ has negligible effect on coordinate distribution of the particle, while it greatly affects the momentum distribution. Therefore, we face a dilemma – if we suppose $b << \sigma_\xi$, then interference picture changes little. However, in this case the correlation between $\xi$ and $x$ will negligibly small. If on the other hand, we suppose $b \sim \sigma_\xi$ or $b >> \sigma_\xi$, then the correlation between $\xi$ and $x$ will be strong and it will be possible to extract information about the particle coordinate $x$ by measuring parameter $\xi$. Nevertheless, the interference picture will be destroyed.

Figure 2 shows disappearance of interference picture due to the interaction between the particle and detector.

So it's impossible to "hook on" anything additional variable $\xi$ which would carry information about initial state, but not change it considerably. Any expansion of quantum system by the way of new variables, which can carry information about its parameters, changes initial quantum state inevitably. The considered property has fundamental value. It characterizes the completeness of quantum statistics (contrary to classical statistics). We shall address this problem again in Section 11.

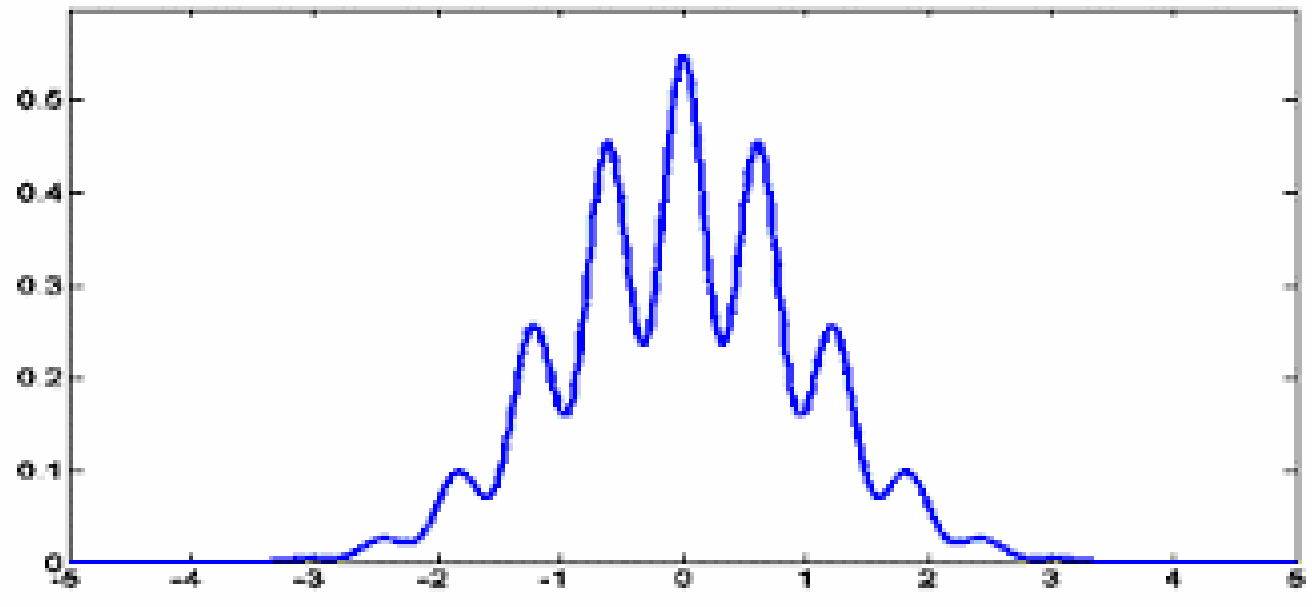

Figure 2. Disappearance of interference picture ($\sigma_x = \sigma_\xi = 0.5 \quad a = 5 \quad b = 0.7$)

**4. Schmidt's modes**

It's remarkable that, in the two-slit interference problem the Schmidt's decomposition for the wave function of system "particle-detector" is given by two modes:

$$\tilde{\psi}(p_x, p_\xi) = \sqrt{\lambda_0}\tilde{\psi}_0^x(p_x)\tilde{\psi}_0^\xi(p_\xi) + \sqrt{\lambda_1}\tilde{\psi}_1^x(p_x)\tilde{\psi}_1^\xi(p_\xi)$$

Zero and first Schmidt's modes are given by:

$$\tilde{\psi}_0^x(p_x) = \left( \frac{2\sqrt{2}\sigma_x}{\sqrt{\pi}\left(1 + \exp\left(-\frac{a^2}{2\sigma_x^2}\right)\right)} \right)^{1/2} \exp\left(-\sigma_x^2 p_x^2\right)\cos(p_x a)$$

$$\tilde{\psi}_1^x(p_x) = \left( \frac{2\sqrt{2}\sigma_x}{\sqrt{\pi}\left(1 - \exp\left(-\frac{a^2}{2\sigma_x^2}\right)\right)} \right)^{1/2} \exp\left(-\sigma_x^2 p_x^2\right)\sin(p_x a)$$

Similarly, for detector the Schmidt's modes are:

$$\tilde{\psi}_0^\xi(p_\xi) = \left( \frac{2\sqrt{2}\sigma_\xi}{\sqrt{\pi}\left(1 + \exp\left(-\frac{b^2}{2\sigma_\xi^2}\right)\right)} \right)^{1/2} \exp\left(-\sigma_\xi^2 p_\xi^2\right)\cos(p_\xi b)$$

$$\tilde{\psi}_1^\xi(p_\xi) = -\left( \frac{2\sqrt{2}\sigma_\xi}{\sqrt{\pi}\left(1 - \exp\left(-\frac{b^2}{2\sigma_\xi^2}\right)\right)} \right)^{1/2} \exp\left(-\sigma_\xi^2 p_\xi^2\right)\sin(p_\xi b)$$

The weights are given by:

$$\lambda_0 = \frac{1+\exp\left(-\frac{a^2}{2\sigma_x^2}\right)+\exp\left(-\frac{b^2}{2\sigma_\xi^2}\right)+\exp\left(-\frac{1}{2}\left(\frac{a^2}{\sigma_x^2}+\frac{b^2}{\sigma_\xi^2}\right)\right)}{2\left[1+\exp\left(-\frac{1}{2}\left(\frac{a^2}{\sigma_x^2}+\frac{b^2}{\sigma_\xi^2}\right)\right)\right]}$$

$$\lambda_1 = \frac{1-\exp\left(-\frac{a^2}{2\sigma_x^2}\right)-\exp\left(-\frac{b^2}{2\sigma_\xi^2}\right)+\exp\left(-\frac{1}{2}\left(\frac{a^2}{\sigma_x^2}+\frac{b^2}{\sigma_\xi^2}\right)\right)}{2\left[1+\exp\left(-\frac{1}{2}\left(\frac{a^2}{\sigma_x^2}+\frac{b^2}{\sigma_\xi^2}\right)\right)\right]}$$

Sum of weights in the Schmidt decomposition is equal to 1:

$\lambda_0 + \lambda_1 = 1$

The interference picture, given by particle's momentum distribution may be represented as the sum of two densities for Schmidt's modes:

$$\tilde{P}_x(p_x) = \lambda_0 \left|\tilde{\psi}_0^x(p_x)\right|^2 + \lambda_1 \left|\tilde{\psi}_1^x(p_x)\right|^2$$

Schmidt modes are illustrated in figure 3.

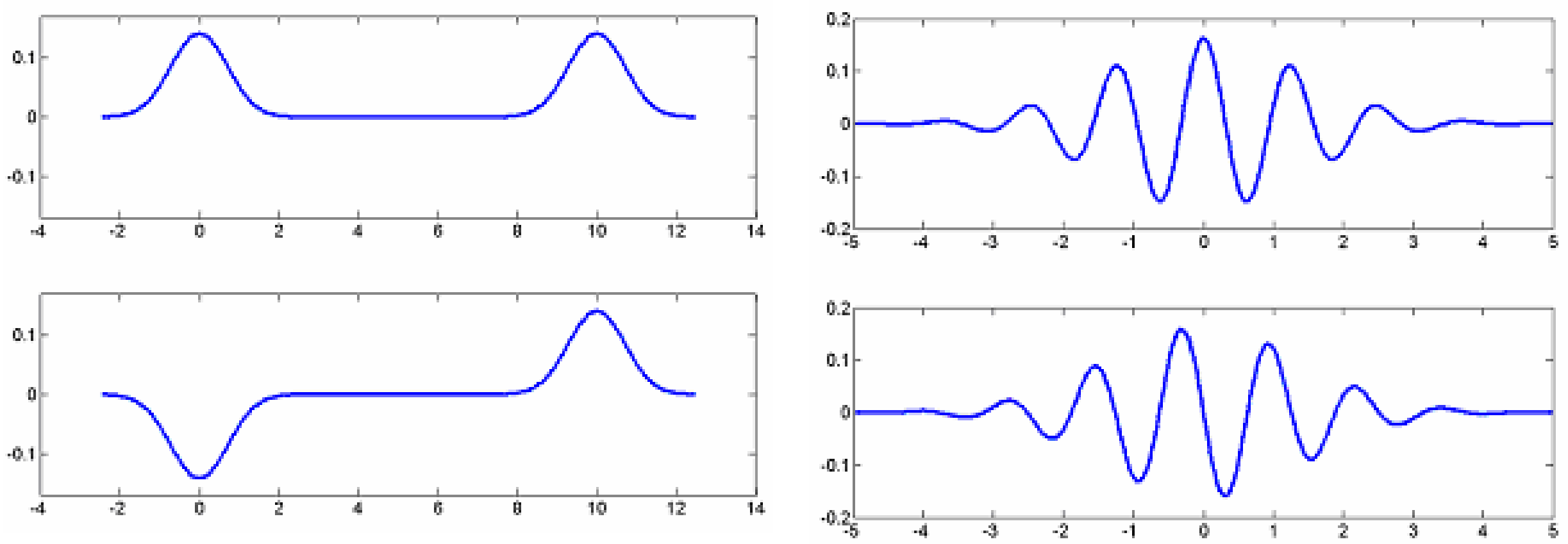

Coordinate modes Momentum modes

Figure 3. Schmidt modes for two-slit interference ($\sigma_x = \sigma_\xi = 0.5$ $a = 5$ $b = 0.5$ $K = 1.4621$ $S = 0.7153$ $\lambda_0 = 0.8033$ $\lambda_1 = 0.1967$)

It's well known that Schmidt's decomposition is closely connected with apparatus of density matrix [1,3]. Density matrices of both subsystems have the same eigenvalues, given by weights in Schmidt decomposition. Furthermore Schmidt's modes of subsystems are eigenvectors of appropriate density matrixes.

The state of a particle considered separately from the detector is given by the following density matrix:

$$\tilde{\rho}_x(p_x, p'_x) = \lambda_0 \tilde{\psi}_0^x(p_x)\tilde{\psi}_0^{x*}(p'_x) + \lambda_1 \tilde{\psi}_1^x(p_x)\tilde{\psi}_1^{x*}(p'_x)$$

**5. Diffraction on a screen with arbitrary number of slits**

Let us consider a flat screen with an arbitrary number of slits $m$ ($m = 2,3,\ldots$).

The probability amplitude of separated ($j$-th) slit entangled with detector is chosen in form:

$$\tilde{\psi}_j(p_x,p_\xi)=\left(\frac{1}{2\pi}\right)^{1/2}\sqrt{2\sigma_x}\sqrt{2\sigma_\xi}\exp\left(-\sigma_x^2 p_x^2\right)\exp\left(-\sigma_\xi^2 p_\xi^2\right)\exp\left(-i\left(p_x x_j^0+p_\xi \xi_j^0\right)\right)$$

Here $x_j^0$ - is the center coordinate of $j$-th slit, $\xi_j^0$ - is the center coordinate of "spot" in detector, arising during transmission of particle through the $j$-th slit.

Let us consider the cases of even and odd numbers of slots separately.

Let $m=2k+1$ - be an odd number. Then the center coordinates of slits and "spot" are accordingly:

$$0,\pm 2a,\pm 4a,\ldots,\pm 2ka$$

$$0,\pm 2b,\pm 4b,\ldots,\pm 2kb$$

Analogously, if $m=2k$ - is an even number then the center coordinates of slits and "spots" are accordingly:

$$\pm a,\pm 3a,\pm 5a,\ldots,\pm(m-1)a$$

$$\pm b,\pm 3b,\pm 5b,\ldots,\pm(m-1)b$$

Here $2a$ - is the distance between consecutive slits $2b$ - is the distance between consecutive "spots".

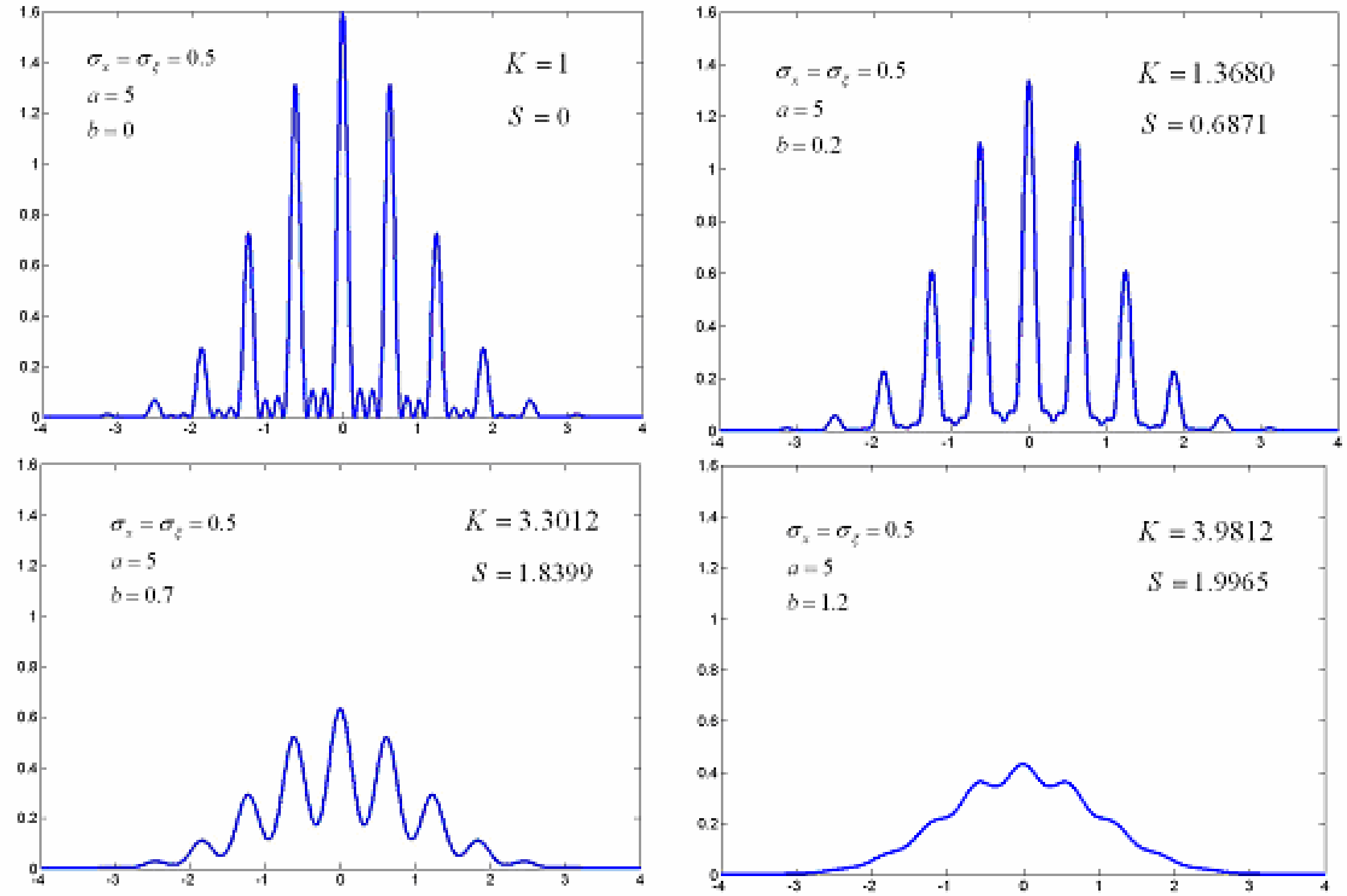

Figure 4. Interference pictures with different levels of interconnection between particle and detector.

The sum probability amplitude from all slits and "spots" in the momentum space is:

$$\tilde{\psi}_j(p_x,p_\xi)=\frac{C}{\sqrt{m}}\left(\frac{1}{2\pi}\right)^{1/2}\sqrt{2\sigma_x}\sqrt{2\sigma_\xi}\exp\left(-\sigma_x^2 p_x^2\right)\exp\left(-\sigma_\xi^2 p_\xi^2\right)F(p_x,p_\xi)$$

Here $C$ - is the normalization factor. In the case of well-defined slits $C \approx 1$,
$F(p_x, p_\xi)$- form- factor which defines entanglement between slits and "spots".
If $m = 2k + 1$ - is an odd number then:

$$F(p_x, p_\xi) = 2\sum_{l=1}^{k} \cos(2\eta l) + 1 = 2\cos((k+1)\eta)\frac{\sin(k\eta)}{\sin(\eta)} + 1,$$

where $\eta = p_x a + p_\xi b$

Analogously if $m = 2k$ - is an even number of slits then:

$$F(p_x, p_\xi) = 2\sum_{l=0}^{k-1} \cos((2l+1)\eta) = 2\cos(k\eta)\frac{\sin(k\eta)}{\sin(\eta)},$$

In the both cases the considered form factors can be expressed by the same formula:

$$F(p_x, p_\xi) = \frac{\sin(m\eta)}{\sin(\eta)},$$

where $\eta = p_x a + p_\xi b$

As an illustration, in the Figure 4, there are interference pictures arising on the screen with four slits with different levels of interconnection between particle and detector.

**6. Informational aspects of diffraction problem**

In a diffraction problem with $m$ slits, the Schmidt decomposition for wave function of system "particle-detector" consists of no more than $m$ components, which have non-zero weights. In one particular case when $b = 0$, there is no entanglement between particle and detector. In this case the state of particle is pure and in the Schmidt's decomposition there is only one (main) mode with weight 1. In the opposite case of $b \neq 0$, the joint state of particle and detector is entangled. In this case Schmidt decomposition has exactly $m$ modes with non-zero weights.

$$\tilde{\psi}(p_x, p_\xi) = \sum_{k=0}^{m-1} \sqrt{\lambda_k}\, \tilde{\psi}_k^x(p_x)\tilde{\psi}_k^\xi(p_\xi)$$

From the informational point of view the interference picture given by density distribution of particle's momentum $\tilde{P}_x(p_x)$, presents a mix of densities of appropriate Schmidt's modes.

$$\tilde{P}_x(p_x) = \sum_{k=0}^{m-1} \lambda_k \left|\tilde{\psi}_k^x(p_x)\right|^2$$

The entropy $S$ is a primary quantity which characterizes quality of interference picture [1-3]:

$$S = -\sum_{k=0}^{m-1} (\lambda_k \log_2 \lambda_k)$$

The others important parameters are the effective number of modes $K$ and information $I$ [13-15].

$$K = \frac{1}{\sum_{k=0}^{m-1} \lambda_k^2}$$

$$I = \log_2 K$$

The entropy $S$ is a quantitative measure of uncertainty of the interference picture. The quality of interference picture is higher for lower $S$. The information $I$, related to Schmidt decomposition characterizes the degree of entanglement between the interfering particle and the detector. The quality of interference picture is higher for lower entanglement and $I$.

In the extreme case of $b \to 0$, only one main mode in Schmidt decomposition ($K \to 1$) has essential value. In this case interference picture has ideal quality, while the entropy $S$ and the information $I$ are close to zero.

In the other extreme case of $b >> \sigma_\xi$, all Schmidt modes have the same weight. In this case $K \to m$, while and entropy and entanglement take their maximum values ($S \to \log_2 m$, $I \to \log_2 m$).

Figure 5 illustrates Schmidt modes for five-slit interference.

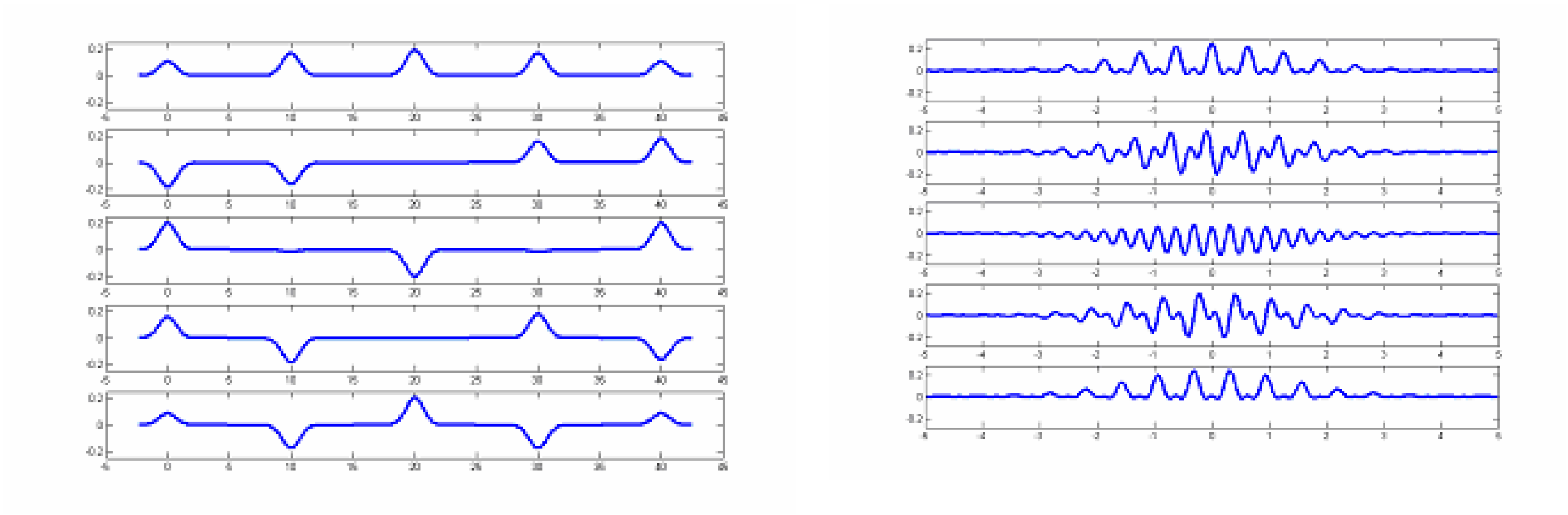

Coordinate modes                Momentum modes

Figure 5 Schmidt modes for five-slit interference:

$\sigma_x = \sigma_\xi = 0.5$, $a = 5$, $b = 0.5$, $K = 3.1043$, $S = 1.8429$;

$\lambda_0 = 0.4434$, $\lambda_1 = 0.3063$, $\lambda_2 = 0.1640$, $\lambda_3 = 0.0666$, $\lambda_4 = 0.0197$

**7. Simulation of coherence loss by means of entanglement with auxilliary quantum system**

The simplest interference scheme considers a plane wave interacting with a screen with two narrow slits. Such picture may be produced if one is put a point source into the focus of thin lens. If we account for finite size of the light source, then the interference picture ceases to be ideal (figure 6).

At first, instead of one point source, let us consider two point sources located at $\pm y$ from the main optical axis. The light beam formed by the source in point $+y$, has angle $\alpha \approx y / F$. As a result, the optical path length corresponding to lower slit is $2a\alpha = 2\varphi$ higher, where $2a$ is the distance between slits and $\varphi = \dfrac{2\pi a y}{\lambda F}$. Here, $2y$ is the distance between point sources, $\lambda$ is the wave length and $F$ is the focal length of lens.

Similarly, we could observe the same pattern for a point source located at $-y$, with the only difference that the upper (not the lower) slit's state lags in phase.

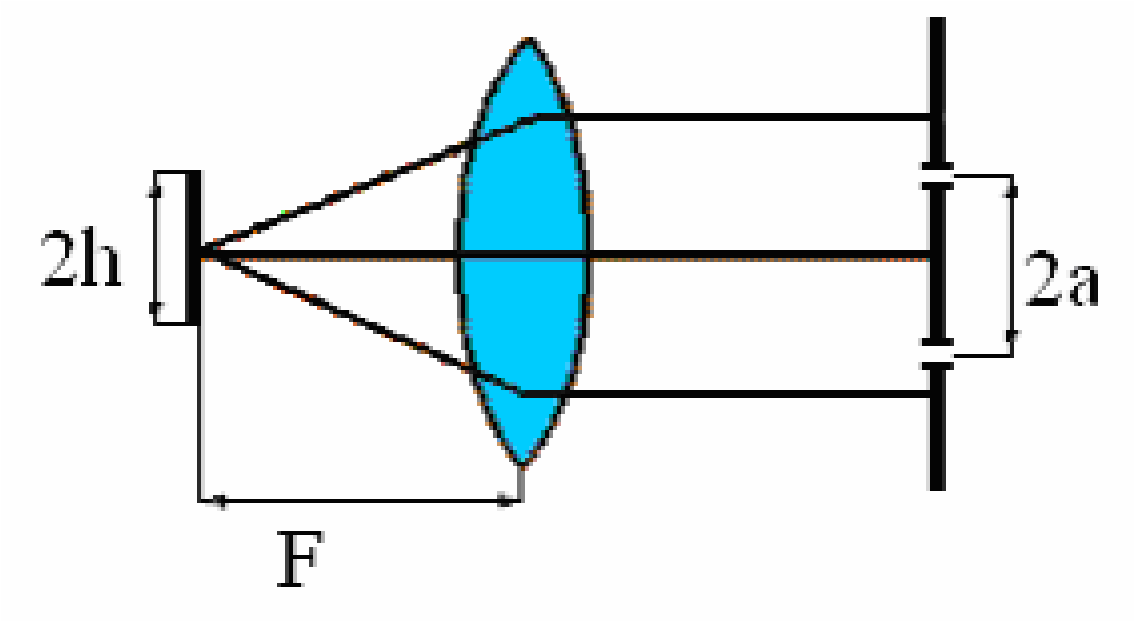

Figure 6. The interference picture from the light source with finite sizes

In the framework of quantum informatics, let us formally consider an entangled state between an interfering particle and an auxilliary qubit (normalization factor is omitted).

$$|\psi\rangle = \left[\exp\left(-\frac{(x-a)^2}{4\sigma_x^2}\right)\exp(-i\varphi) + \exp\left(-\frac{(x+a)^2}{4\sigma_x^2}\right)\exp(i\varphi)\right]|0\rangle +$$
$$\left[\exp\left(-\frac{(x-a)^2}{4\sigma_x^2}\right)\exp(i\varphi) + \exp\left(-\frac{(x+a)^2}{4\sigma_x^2}\right)\exp(-i\varphi)\right]|1\rangle$$

During measurement, the auxilliary qubit is registered in state $|0\rangle$ with probability 0.5, and the interfering particle is then registered in state $\left[\exp\left(-\frac{(x-a)^2}{4\sigma_x^2}\right)\exp(-i\varphi) + \exp\left(-\frac{(x+a)^2}{4\sigma_x^2}\right)\exp(i\varphi)\right]$. Similarly, the qubit is registered in state $|1\rangle$ with probability 0.5, while the interfering particle is in state $\left[\exp\left(-\frac{(x-a)^2}{4\sigma_x^2}\right)\exp(i\varphi) + \exp\left(-\frac{(x+a)^2}{4\sigma_x^2}\right)\exp(-i\varphi)\right]$.

The main characteristic of the quality of a two-slit interference picture is so-called visibility. It's defined by the following formula [17, 18]:

$$V = \frac{I_{\max} - I_{\min}}{I_{\max} + I_{\min}}$$

where $I_{\max}$ and $I_{\min}$ are maximum and minimum intensities correspondingly. Here it is also supposed that sizes of slits are much smaller than the distance between them.

Methodically, a quantum information interpretation of spatial optical coherence is obtained by enlarging the classical picture to a quantum state.

On the one hand, an interference picture may be described by its visibility parameter $V$, while on the other hand it is a function of Schmidt number $K$. The interfering picture may be considered on the one hand as a function of parameter visibility, and on the other hand as a function of Schmidt's number. The comparison of these ways leads to universal coupling between visibility and Schmidt number:

$$V = \sqrt{\frac{2-K}{K}} \quad \text{or} \quad K = \frac{2}{1+V^2}$$

Graphically the coupling between visibility and Schmidt's number is represented on the figure 7.

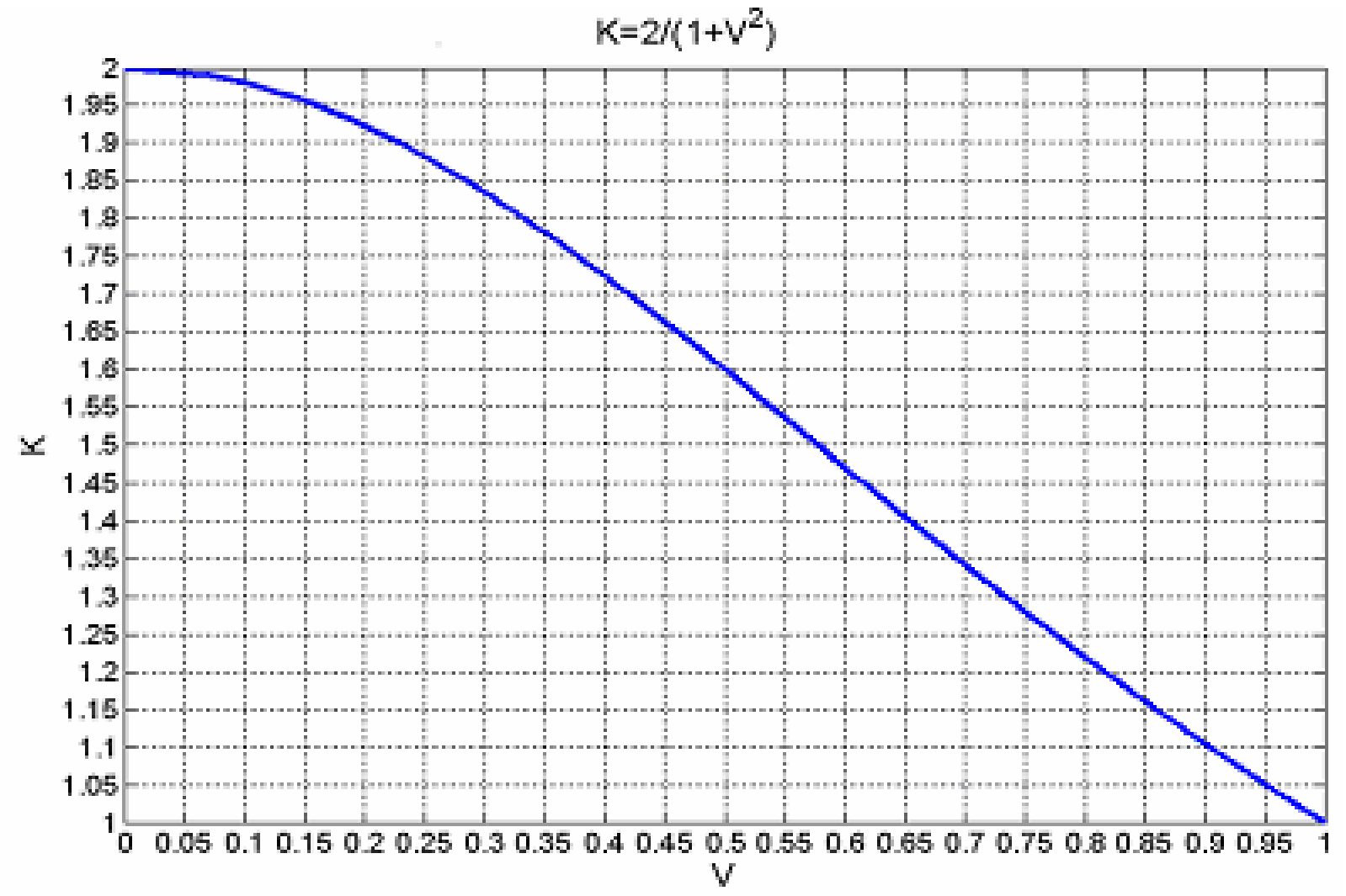

Figure 7. Coupling between visibility and Schmidt's number

Let's consider the source with finite sizes. Let $2h$ - is a size of the source, $2a$ - is a distance between slits

$$y = \frac{ah}{\lambda F}$$ - is non-dimensional parameter of source's size.

The visibility dependence on source's size is given by formula [17, 18]:

$$V(y) = \left|\operatorname{sinc}(4y)\right|$$

where
$$\operatorname{sinc}(x) = \begin{cases} \dfrac{\sin(\pi x)}{\pi x}, & x \neq 0 \\ 1, & x = 0 \end{cases}$$

Well, due to finding of dependence above the Schmidt's number as a function of source will be given by formula:

$$K(y) = \frac{2}{1+\operatorname{sinc}^2(4y)}$$

The represented above analytical results are confirmed by numerical calculations.

The coupling between visibility and entanglement is very important. It shows numerically how the information connection between quantum particle and its environment leads to degradation of interfering picture. This question from some different positions also will be discussed in the section 11.

In the case of two-slit interference there are only two Schmidt modes and the sum of their weights equals to 1. So all possible interference pictures can be described as one-parameter family. In the capacity of parameter it can be been for example visibility $V$, the main Schmidt mode weight $\lambda_0$, Schmidt number $K$, and also entropy $S$. The coupling between visibility and Schmidt weights is defined by formulas:

$$\lambda_0 = \frac{1+V}{2}, \quad \lambda_1 = \frac{1-V}{2}$$

The coupling between visibility and entropy is given by the expression:

$$S = -\left[\frac{1+V}{2}\log_2\left(\frac{1+V}{2}\right) + \frac{1-V}{2}\log_2\left(\frac{1-V}{2}\right)\right]$$

The obtaining formula allows determine visibility degrading degree for interference picture subject to entropy level which is introduced during the interaction of particle with environment.

**8. The particle in the double- well potential with tunnel coupling**

Let's consider the movement of quantum particle in double- well potential with tunnel coupling (figure 8).

$$U(x) = -\frac{\alpha x^2}{2} + \frac{\beta x^4}{4}, \quad \alpha > 0, \ \beta > 0$$

Note that this potential may be considered as model basis for qubit on quantum dot separated by controled potential barrier [19]. Furthermore as it'll be showed in the section 10 the represented potential is very useful for considering an ammonia molecule as a qubit [20].

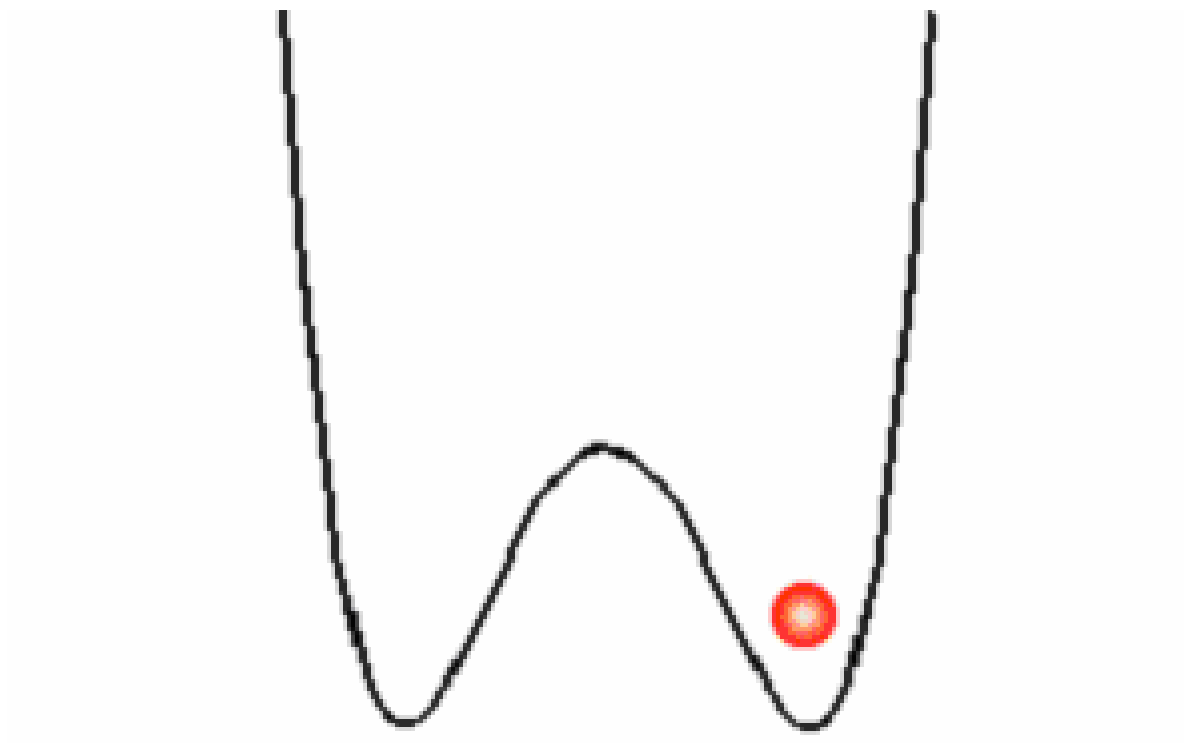

Figure 8. Quantum particle in double- well potential with tunnel coupling

The possible points of equilibrium are found from the condition:

$$\frac{\partial U}{\partial x} = 0$$

The point $x = 0$ corresponds to unstable equilibrium and points $x = \pm a$, where $a = \sqrt{\frac{\alpha}{\beta}}$ -to stable.

The minimum value of potential energy in the equilibrium points is:

$$U_{\min} = -\frac{\alpha^2}{4\beta}$$

Near the points of stable equilibrium it's possible harmonic oscillations of particle. The rigidity of appropriate "springs" equals to $\frac{\partial^2 U}{\partial x^2}\Big|_{x=\pm a} = 2\alpha$. The frequency of free oscillations is: $\omega_0 = \sqrt{\frac{2\alpha}{m}}$

The lowest quantum state amassing near the bottom of potential well is approximately the basis state of oscillator

$$\psi_0(x) = \left(\frac{m\omega_0}{\pi}\right)^{1/4} \exp\left(-\frac{m\omega_0 (x - x_0)^2}{2}\right)$$

The considered state is a Gauss state. The appropriate dispersion for it is:

$$\sigma_x^2 = \frac{1}{2m\omega_0}$$

The basis and excited states of particle in the considered double- well potential with tunnel coupling will be approximately given by symmetric and antisymmetric wave functions accordingly. For the symmetric basis state there is:

$$\psi_0(x) = C_0\left(\frac{1}{\sqrt{2\pi}\,2\sigma_x}\right)^{1/2}\left[\exp\left(-\frac{(x-a)^2}{4\sigma_x^2}\right) + \exp\left(-\frac{(x+a)^2}{4\sigma_x^2}\right)\right],$$

where $C_0$ - is normalization constant:

$$C_0^2 = \frac{1}{1 + \exp\left(-\frac{a^2}{2\sigma_x^2}\right)}$$

Analogously for the antisymmetric combination there is:

$$C_1^2 = \frac{1}{1 - \exp\left(-\frac{a^2}{2\sigma_x^2}\right)}$$

In the momentum representation the considered states are given by formulas:

$$\tilde{\psi}_0(p_x) = 2C_0\left(\frac{1}{2\pi}\right)^{1/4}\sqrt{\sigma_x}\,\exp\left(-\sigma_x^2 p_x^2\right)\cos(p_x a)$$

$$\tilde{\psi}_1(p_x) = 2iC_1\left(\frac{1}{2\pi}\right)^{1/4}\sqrt{\sigma_x}\,\exp\left(-\sigma_x^2 p_x^2\right)\sin(p_x a)$$

Note that represented formulas are corresponded to formulas for two-slit interference (see the section 2).

It's more comfortable to make a calculation for kinetic energy of ground and excited states in the momentum representation. The result is the next:

$$T_0 = \frac{C_0^2}{8m\sigma_x^2}\left[1-\left(\frac{a^2}{\sigma_x^2}-1\right)\exp\left(-\frac{a^2}{2\sigma_x^2}\right)\right]$$

$$T_1 = \frac{C_1^2}{8m\sigma_x^2}\left[1+\left(\frac{a^2}{\sigma_x^2}-1\right)\exp\left(-\frac{a^2}{2\sigma_x^2}\right)\right]$$

Analogously potential energy of ground and excited states is (here it's comfortable to use coordinate representation):

$$U_0 = C_0^2\left[-\frac{\alpha}{2}\left(a^2+\sigma_x^2\right)+\frac{\beta}{4}\left(a^4+6\sigma_x^2a^2+3\sigma_x^4\right)+\exp\left(-\frac{a^2}{2\sigma_x^2}\right)\left(-\frac{\alpha\sigma_x^2}{2}+\frac{3\beta\sigma_x^4}{4}\right)\right]$$

$$U_1 = C_1^2\left[-\frac{\alpha}{2}\left(a^2+\sigma_x^2\right)+\frac{\beta}{4}\left(a^4+6\sigma_x^2a^2+3\sigma_x^4\right)-\exp\left(-\frac{a^2}{2\sigma_x^2}\right)\left(-\frac{\alpha\sigma_x^2}{2}+\frac{3\beta\sigma_x^4}{4}\right)\right]$$

So the full energy of basis and excited states is:

$$E_0 = U_0 + T_0$$

$$E_1 = U_1 + T_1$$

The considered analysis is based on supposing that it's possible to limit by linear combinations of basis states in wells and not consider excited oscillator's states. It leads to the condition:

$$\exp\left(-\frac{a^2}{2\sigma_x^2}\right) << 1$$

**9. Calculation of qubit interaction**

Let there be two particles moving in the double- well potentials with tunnel coupling as described above in section 8. To the approximation of non-interactive particles there are four states which make basis states:

$$\psi_{00}(x,\xi) = \psi_0(x)\psi_0(\xi)$$

$$\psi_{01}(x,\xi) = \psi_0(x)\psi_1(\xi)$$

$$\psi_{10}(x,\xi) = \psi_1(x)\psi_0(\xi)$$

$$\psi_{11}(x,\xi) = \psi_1(x)\psi_1(\xi)$$

In the absence of interaction the Hamilton matrix of two-qubit system has a diagonal form:

$$H_0 = \begin{pmatrix} 2E_0 & 0 & 0 & 0 \\ 0 & E_0+E_1 & 0 & 0 \\ 0 & 0 & E_0+E_1 & 0 \\ 0 & 0 & 0 & 2E_1 \end{pmatrix}$$

Let the form of Hamiltonian's interaction between qubits have the delta-function form:

$$H_{\text{int}} = -g_0\delta(x-\xi)$$

The case $g_0 > 0$ corresponds to attraction of particles and $g_0 < 0$ to their repulsion.

Note that this potential was used early by D.I. Blokhintsev for considering the process of quantum measurements [21].
Turning on interaction leads to that Hamiltonian matrix becoming non-diagonal:

$$H = H_0 + h,$$

where $h$ - is the excitement factor, appearing during to the interaction:

$$h = \begin{pmatrix} h_1 & 0 & 0 & h_2 \\ 0 & h_2 & h_2 & 0 \\ 0 & h_2 & h_2 & 0 \\ h_2 & 0 & 0 & h_3 \end{pmatrix}$$

Here:

$$h_1 = h_{11} = -\frac{C_0^4 g_0}{4\sqrt{\pi}\sigma_x}\left(1 + 4\exp\left(-\frac{3a^2}{4\sigma_x^2}\right) + 3\exp\left(-\frac{a^2}{\sigma_x^2}\right)\right)$$

$$h_2 = h_{14} = -\frac{C_0^2 C_1^2 g_0}{4\sqrt{\pi}\sigma_x}\left(1 - \exp\left(-\frac{a^2}{\sigma_x^2}\right)\right)$$

$$h_3 = h_{44} = -\frac{C_1^4 g_0}{4\sqrt{\pi}\sigma_x}\left(1 - 4\exp\left(-\frac{3a^2}{4\sigma_x^2}\right) + 3\exp\left(-\frac{a^2}{\sigma_x^2}\right)\right)$$

Note that in approximation

$$h_1 \approx h_2 \approx h_3 \approx -\frac{g_0}{4\sqrt{\pi}\sigma_x}$$

Stationary states of the system of two interacting qubits are given by eigenvectors of matrix $H$, and the energies of these states are given by its eigenvalues.

Let the two qubit state have the following form:

$$\psi(x,\xi) = c_{00}\psi_{00}(x,\xi) + c_{01}\psi_{01}(x,\xi) + c_{10}\psi_{10}(x,\xi) + c_{11}\psi_{11}(x,\xi)$$

It can be shown that in this case, the Schmidt number, characterizing the degree of entanglement of qubits is:

$$K = \frac{1}{1-2\Delta},$$

where: $\Delta = \left|c_{00}c_{11} - c_{01}c_{10}\right|^2$

**10. Application for ammonia molecule**

The considered above model can be applied to the classical example of molecule of ammonia $NH_3$, which is a right regular pyramid (figure 9). Its base consists of three atoms of hydrogen $H$, and the top point is atom of nitrogen $N$. The angles between bonds $H-N-H$ equal $108^\circ$, and interatomic distances are $1.015\,\mathring{A}$ for lateral edges $N-H$, made by nitrogen and hydrogen atoms and $1.64\,\mathring{A}$ for sides $H-H$ of triangle made by hydrogen in the base of pyramid (numerical data is taken from the third issue of Big Soviet Encyclopedia).

Due to tunnel effect ammonia molecule can "turn inside out" (so-called structure inverse). It can be presented this way: the plane of pyramid base made by three atoms of hydrogen is either at the left or at the right from the atom of nitrogen (the molecule axis is considered as horizontal). The difference of energies between antisymmetric and symmetric combinations of ammonia molecule equals to 24GHz – the classical transition frequency underlying in the laser of Basov and Prokhorov. Due to practical importance of this transition it is often discussed in literature [22, 23]. As it has been already noted, there are some proposals to realize a quantum computer based on ammonia molecules put into fullerene molecules [20].

$NH_3$

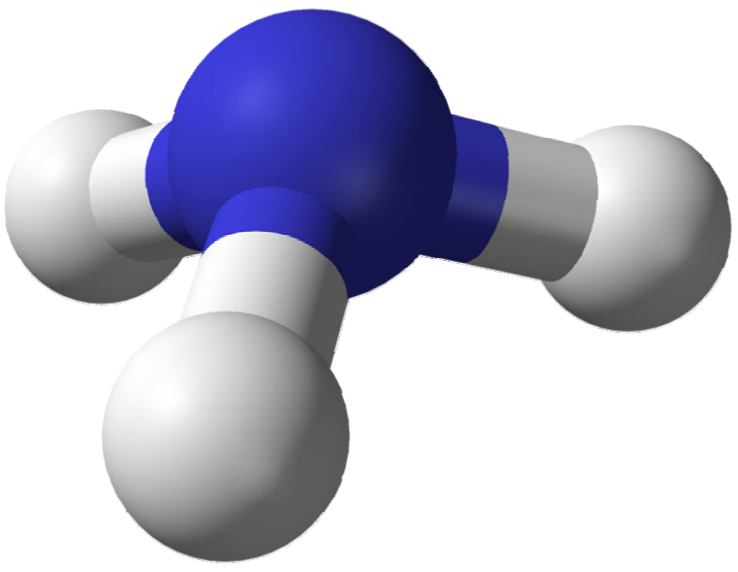

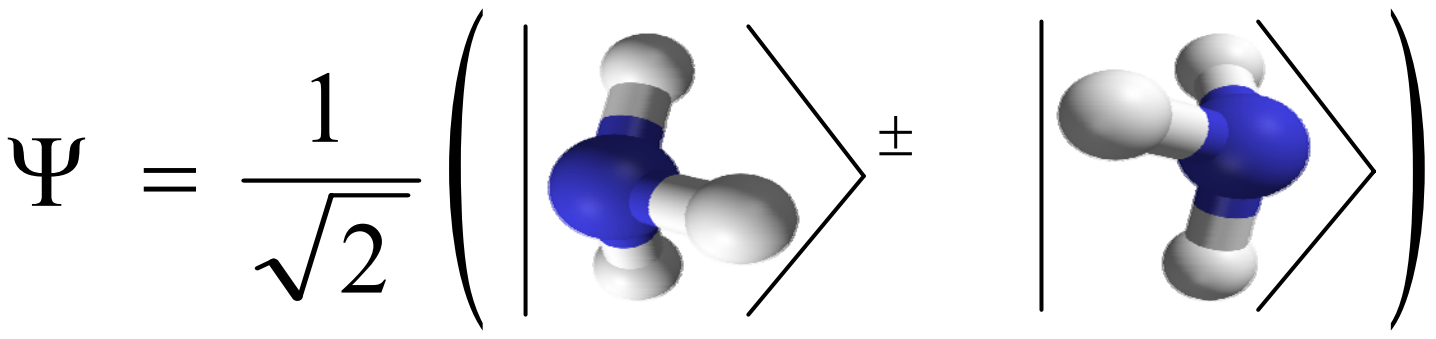

Figure 9. Molecule of ammonia and its quantum state

Let us present the considered process of tunnel inverse as one-dimensional two-particle interaction: one of them is atom of nitrogen and the other particle consists of three atoms of hydrogen. So base of known transition's frequency 24GHz in molecule $NH_3$ we can estimate parameters of potential $\alpha$ and $\beta$. For this purpose we need to note that reduced mass for three hydrogen atoms relative to nitrogen atom equals $m = \frac{3 \cdot 14}{3+14} = 2.47$ atomic mass unit and height of pyramid is an equilibrium distance between nitrogen and hydrogen plane, and as it can be seen from simple geometric considerations it equals $a = 0.37\ \mathring{A}$.

Let us use non-dimensional system of units in which $\hbar = 1$, mass of particles is measured in atomic mass units $m_0 = 1.66 \cdot 10^{-27}$ кг, distance is given in Bohr radius $a_0 = 0.529\ \mathring{A}$. In Bohr radius units the equilibrium positions of effective particle equals $x_0 = \pm a$, where $a = 0.699$. In this case a unit of energy has value $\frac{\hbar^2}{m_0 a_0^2} = 2.39 \cdot 10^{-21}$ Дж. In this units the transition energies 24GHz correspond to energy difference between excited and basis states equal to $E_1 - E_0 = 0.00665$. The selection of parameters $\alpha$ and $\beta$, providing right values for $a$ and $E_1 - E_0$ leads to the next result: $\alpha = 69.29$ $\beta = 141.73$.

Note that the condition of applying perturbation theory is performed with good precision i.e. $\exp\left(-\frac{a^2}{2\sigma_x^2}\right) = 0.000118 << 1$

It is interesting that offered theory admits empirical testing. The fact of the matter is that along with usual form of ammonia $NH_3$, there is its isotopic modification $ND_3$, in which the hydrogen $H$ is replaced by deuterium $D$. Let us change the reduced mass on $m = \frac{6 \cdot 14}{6+14} = 4.2$ (it will correspond to $ND_3$) without changes to parameters $\alpha$ and $\beta$. So by direct estimation we will get $E_1 - E_0 = 0.000416 = 1.5\ \text{ГГц}$. The prediction corresponds to changing transition frequency from 24GHz to 1.5GHz when $NH_3$ is replaced by $ND_3$. The considered effect for sharp decreasing of transition energy is due to more light tunneling in molecule $NH_3$, in comparison with $ND_3$: $\exp\left(-\frac{a^2}{2\sigma_x^2}\right) = 7.5 \cdot 10^{-6}$

The experimental value of transition frequency for $ND_3$ equals to 1.6 GHz [24], and it corresponds to theoretical estimation mentioned above. So the model works very well despite its simplicity and its one-dimensional form.

It should be noted that along with $NH_3$ and $ND_3$, there is one more isotopic ammonia modification - $NT_3$, in which hydrogen atoms are replaced by atoms of tritium. Now the reduced mass of particle equals $m = \frac{9 \cdot 14}{9+14} = 5.48$. The considered model gets the value of 0,28 GHz for transition frequency and experiment gets 0,306 GHz [25]. Thus the experiment is again in good agreement with the theory.

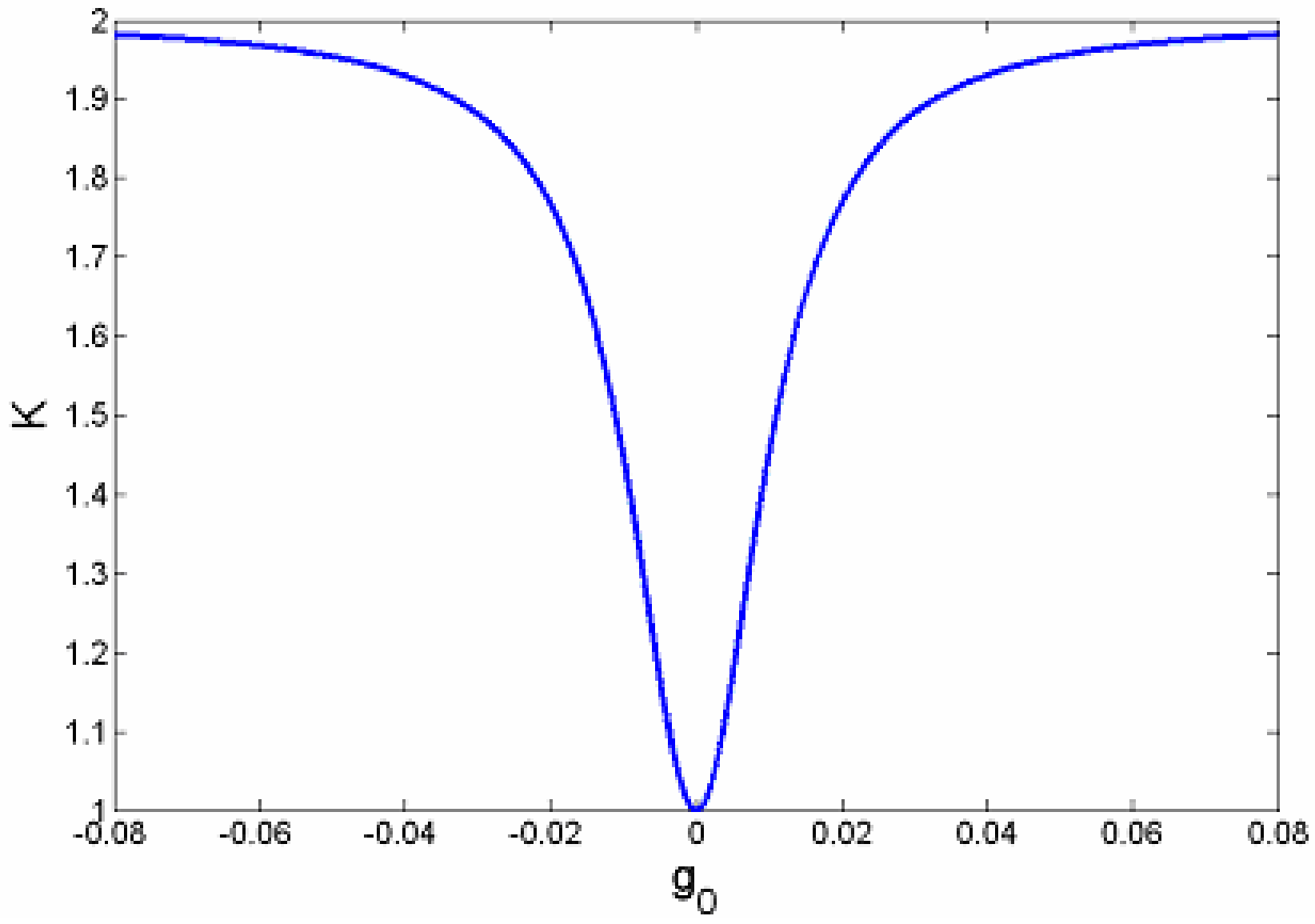

Figure 10. Dependence of degree of entanglement on parameters of intensity in a two-qubit system of ammonia molecule.

In the liquid phase between nitrogen molecules there are hydrogen bonds which in the considered case may be explained by dipole-dipole interaction. The hydrogen bonds fix space placement of pyramids and deny to "turn inside out" (it corresponds to considered two-qubit model). The fixation of bonds leads to disappearing of tunnel inverse transition in liquid. In fact the considered transition decreases not only in liquid but in gas under pressure as well [24].

Figure 10 shows dependence of degree of entanglement K on intensity of interaction $g_0$ for ground state of a two-qubit system based on ammonia molecule. In dimensionless parameters of present section $E_1 - E_0 = 0.00665$, $\sigma_x = 0.164$

Numerical calculations show that, increasing the interaction force leads to increases in entanglement - from $K = 1$ when $g_0 = 0$ to $K \to 2$ when $\frac{|g_0|}{4\sqrt{\pi}\sigma_x} >> E_1 - E_0$

This analysis provides arguments for the benefits of interference nature of ammonia molecule quantum state. Observing this state coupled with interaction between molecule and environment leads to destroying this interfering state in the same analogy with experiments of the "which way" type.

**11. Adequacy and completeness of measurements**

The purpose of this section is to obtain quantative limits for interaction levels between a quantum system and its environment. These limits follow from results of quantum measurements.

Let us use matrix form for our calculations. For this purpose "stretch" the density matrix $\rho$ into column (take the second column and put it under the first and so on). If matrix $B$ is some protocol measurement matrix, then the results of quantum measurements look like a system of linear equations:

$$B\rho = P,$$

where $P$ - are measurement probabilities (probability density).

Let us take $n$ points from each space (coordinate and momentum). Then $P$ will be a column of length $2n$ ($N = 2n$).

The density matrix consists of $s^2$ elements where $s$ - is the dimension of Hilbert space. Accordingly the matrix $B$ has dimension $2n \times s^2$.

The matrix $B$ is represented in so-called SVD-decomposition (Singular Value Decomposition) form.

$$B = USV^+,$$

where $U$ and $V$ - are unitary matrices and $S$ - is diagonal positive matrix (its diagonal elements are named singular values). The sizes of considered matrixes are equal to the following values: for $U$ - $2n \times 2n$, for $S$ - $2n \times s^2$, for $V$ - $s^2 \times s^2$.

Let us introduce new variables. Instead of the independent variable (column $\rho$) we shall consider a unitary coupled with it variable, which we shall call the column of factors $f$:

$$f = V^+\rho.$$

Similarly, instead of initial measurements (column $P$ ), we shall consider a unitary coupled with it column $Q$, which will be named the characteristic column.

$$Q = U^+ P$$

In new notations the considered system is reduced to the following form:

$$Sf = Q$$

This system is elementary because $S$ is a diagonal matrix. Its analysis allows us to classify measurements from the viewpoint of adequacy and completeness.

Let $2n > s^2$ ($N > s^2$),i.e. the number of measurements is greater than the number of density matrix elements (it's a typical situation). The rank of model $r$ it is the number of non-zero singular values. Evidently, $r \le s^2$. In matrix $S$ the last $2n - r$ strings are knowingly equal to zero. Hence it means that for consistency of the system, the last $2n - r$ ($N - r$) values in the characteristic column should be equal to zero. Let us call this condition - the condition of **adequacy** for measurements. If this condition is not met, then there is inadequacy: statistical distributions knowingly do not correspond to the considered quantum mechanics model. That may imply that either the experiment was done incorrectly (or numerical calculations), or not all basis functions were taken into account or this functions were chosen incorrectly. From the mathematical viewpoint inadequacy leads to the impossibility to solve the system of equations, so that it is not consistent.

Let the system meet the adequacy condition. If all singular values are knowingly different from zero i.e. $r = s^2$, then there is **unconditional completeness** and the solution exists. The measurement protocol completely defines any quantum state (as pure as mixed) in the considered Hilbert space.

In this case we shall get the factor column dividing the elements of the characteristic column by corresponding singular values.

$$f_j = Q_j / S_j \quad j = 1,2,...,s^2$$

As a result, the density matrix is derived by a unitary transformation:

$$\rho = Vf$$

Now let us consider the case of $r < s^2$, i.e. some of singular values are equal to zero. Then for non-zero singular values we have:

$$f_j = Q_j / S_j \qquad j = 1,2,...,r$$

We shall call these factors $f_j$ $j = 1,2,...,r$ as defined factors.

At the same time for zero singular values we have equations corresponding to the uncertainty of type "zero divided by zero".

$$0 f_j = 0 \qquad\qquad j = r+1, r+2,...,s^2$$

We shall call these factors $f_j$ $j = r+1, r+2,...,s^2$ as undefined factors.

We can take any arbitrary complex numbers as the solutions the latter equations. The considered situation corresponds to the case of incompleteness of measurements. The system of equations has an infinite set of solutions. Hence not all of them correspond to physical density matrices. The physical solutions are only those ones that give a Hermitian and non-negatively defined density matrix. Formally, all such solutions can be obtained by "scanning" all possible values of undefined factors. It's understandable that such procedure can be realistically implemented only if the space dimension of undefined factors is not too high.

**The regularized (normal)** solution shall correspond to the choice of zero values as the undefined factors: $f_j = 0 \quad j = r+1, r+2, ..., s^2$. Due to the unitary property of the interdependence between the density matrix and the factor vector, the regularized solution gives the following top-boundary **limit for Schmidt's number**:

$$K \le K_{\max}\text{, где } K_{\max} = 1/\left(f^+ f\right).$$

This limit has an important significance - due to incompleteness of measurement protocol we can not know the exact solution and the procedure of finding all solutions may be very complex. However knowledge of $K_{\max}$, in any case gives us confidence that the level of correlation (entanglement) $K$ of our quantum system with its environment can not exceed this value and must fall into the interval $1 \le K \le K_{\max}$.

The result corresponding to $K_{\max} = 1$ is especially remarkable. In this case it is possible to draw the conclusion that $K = 1$. It means that the considered system can not be entangled with another system. Therefore, we have complete information about our quantum state despite our incomplete set of measurements. The protocol corresponding to $K_{\max} = 1$, will be named **conditionally complete**: the completeness property is met for some specially selected pure states.

The interfering quantum states considered above in sections 2, 5 and 8 correspond to the case of conditional completeness. Because $K_{\max} = 1$, there is no possibility to get any information about our considered system by the way of entangling it with any detector. Any such entanglement leads to $K > 1$, which alters the initial results of measurements (blooming of interference picture).

## 12. Conclusion

Let us briefly formulate the main results:

- A simple analytical model which describes the joint entangled state of interfering particle and detector is offered. The considered model allows one to give a visual quantitative description for the process of interference picture degradation with growing interaction between the particle and the detector.
- An analysis of the problem "which way" using Schmidt mode methodology is given. Explicit expressions for particle and detector modes, as well as Schmidt weights are given. It is shown that the interference picture given by density distribution of the particle momentum is a mix of densities of corresponding Schmidt modes.
- For the first time, a quantum information approach has been developed to describe classical coherence. In this framework, the interconnection between visibility of interference picture and Schmidt number is studied.
- The problem of quantum particle tunneling between two potential wells is considered. The model is applied to practical description of ammonia molecule and its isotopic modifications. Good correspondence between theory and experiment is observed.
- It's shown that quantum measurements impose some limits on the possible form of quantum state. Quantitatively, this restriction is reduced to determining the maximum possible Schmidt number, which characterizes the degree of entanglement of the system with its environment.

**Acknowledgments**

Useful discussions with Academician K.A. Valiev, Professor S. P. Kulik and Professor V. V. Vyurkov are gratefully acknowledged.